% Context-Free Multilanguages
\magnification\magstephalf
\def\capdot{\mathchoice{\cpdt{}}{\cpdt{}}{\cpdt\scriptstyle}%
 {\cpdt\scriptscriptstyle}}

\def\cpdt{\gcpdt\mathbin\cap.}
\def\bcpdt{\gcpdt\mathop\bigcap{\lower.3\ht0\hbox{\bf.}}}
\def\gcpdt#1#2#3#4{#1{\setbox0=\hbox{$#4#2$}\vtop{\copy0
  \vskip-\baselineskip\kern-.2\ht0\hbox to\wd0{\hss#3\hss}}}}
\def\bn{\bigskip\noindent}
\def\Piit{{\mit\Pi}}
\def\Sigmait{{\mit\Sigma}}
\def\ra{\rightarrow}
\def\disleft#1:#2:#3\par{\par\hangindent#1\noindent
	 \hbox to #1{#2 \hfill \hskip .1em}\ignorespaces#3\par}

\hyphenation{multi-language multi-languages}

\centerline{\bf Context-Free Multilanguages}
\centerline{Donald E. Knuth}
\centerline{Computer Science Department, Stanford University}

\bn
Inspired by ideas of Chomsky, Bar-Hillel, Ginsburg, and their
coworkers, I~spent the summer of 1964 drafting Chapter~11 of a book
I~had been asked to write. The main purpose of that book, tentatively
entitled {\sl The Art of Computer Programming}, was to explain how to
write compilers; compilation was to be the subject of the twelfth and
final chapter. Chapter~10 was called ``Parsing,'' and Chapter~11 was
``The theory of languages.'' I~wrote the drafts of these chapters in
the order 11, 10, 12, because Chapter~11 was the most fun to do.

Terminology and notation for formal linguistics were in a great state
of flux in the early~60s, so it was natural for me to experiment with
new ways to define the notion of what was then being called a ``Chomsky
type~2'' or ``ALGOL-like'' or ``definable'' or ``phrase structure'' or
``context-free'' language. As I~wrote Chapter~11, I~made two changes
to the definitions that had been appearing in the literature. The first
of these was comparatively trivial, although it simplified the
statements and proofs of quite a few theorems: I~replaced the
``starting symbol''~$S$ by a ``starting set'' of strings from which
the language was derived. The second change was more substantial:
I~decided to keep track of the multiplicity of strings in the language,
so that a string would appear several times if there were several ways
to parse~it. This second change was natural from a programmer's
viewpoint, because transformations on context-free grammars had proved
to be most interesting in practice when they yielded isomorphisms
between parse trees.

I never discussed these ideas in journal articles at the time, because
I~thought my book would soon be ready for publication. (I~published an
article about LR$(k)$ grammars~[4]
only because it was an idea that occurred to me after finishing the
draft of Chapter~10; the whole concept of LR$(k)$ ws well beyond the
scope of my book, as envisioned in 1964.) My paper on parenthesis
grammars~[5]
did make use of starting sets, but in my other relevant papers~[4, 6,~8]
I~stuck with the more conventional use of a starting symbol~$S$.
I~hinted at the importance of multiplicity in the answer to exercise
4.6.3--19 of {\sl The Art of Computer Programming\/} (written in 1967,
published in 1969~[7]):
``The terminal strings of a noncircular context-free grammar form a
multiset which is a set if and only if the grammar is unambiguous.''
But as the years went by and computer science continued its explosive
growth, I~found it more and more difficult to complete final drafts
of the early chapters, and the date for the publication of Chapter~11
kept advancing faster than the clock was ticking.

Some of the early literature of context-free grammars referred to
``strong equivalence,'' which meant that the multiplicities 
0, 1, and $\geq 2$
were preserved; if ${\cal G}_1$
was strongly equivalent to~${\cal G}_2$, then ${\cal G}_1$ was
ambiguous iff ${\cal G}_2$ was ambiguous. But this concept did not
become prominent enough to deserve mention in the standard textbook on
the subject~[1].

The occasion of Seymour Ginsburg's 64th birthday has reminded me that
the simple ideas I~played with in~`64 ought to be aired before too
many more years go~by. Therefore I~would like to sketch here the basic
principles I~plan to expound in Chapter~11 of {\sl The Art of Computer
Programming\/} when it is finally completed and published---currently
scheduled for the year 2008. My treatment will be largely informal,
but I~trust that interested readers will see easily how to make
everything rigorous. If these ideas have any merit they may lead some
readers to discover new results that will cause further delays in the
publication of Chapter~11. That is a risk I'm willing to take.

\bn{\bf 1. Multisets.}\enspace
A {\it multiset\/} is like a set, but its elements can appear more
than once. An element can in fact appear infinitely often, in an
infinite multiset. The multiset containing 3~$a$'s and 2~$b$'s can be
written in various ways, such as $\{a,a,a,b,b\}$, $\{a,a,b,a,b\}$, or
$\{3\cdot a,\,2\cdot b\}$. If $A$ is a multiset of objects and if $x$
is an object, $[x]\,A$ denotes the number of times $x$ occurs in~$A$;
this is either a nonnegative integer or~$\infty$. We have $A\subseteq
B$ when $[x]\,A\leq [x]\,B$ for all~$x$; thus $A=B$ if and only
$A\subseteq B$ and $B\subseteq A$. A~multiset is a {\it set\/} if no
element occurs more than once, i.e., if $[x]\,A\leq 1$ for all~$x$. If
$A$ and~$B$ are multisets, we define $A^{\cap}$, $A\cup B$, $A\cap B$,
$A\uplus B$, and $A\capdot B$ by the rules
$$\eqalign{[x]\,A^{\cap}&=\min(1,[x])\,;\cr
[x]\,(A\cup B)&=\max([x]\,A,\,[x]\,B)\,;\cr
[x]\,(A\cap B)&=\min([x]\,A,\,[x]\,B)\,;\cr
[x]\,(A\uplus B)&=([x]\,A)+([x]\,B)\,;\cr
[x]\,(A\capdot B)&=([x]\,A)+([x]\,B)\,.\cr}$$
(We assume here that $\infty$ plus anything is $\infty$ and that
0~times anything is~0.)
Two multisets $A$ and~$B$ are {\it similar}, written $A\asymp B$, if
$A^{\cap}=B^{\cap}$; this means they would agree as sets, if
multiplicities were ignored. Notice that $A\cup B\asymp A\uplus B$ and
$A\cap B\asymp A\capdot B$. All four binary operations are associative
and commutative; several distributive laws also hold, e.g.,
$$(A\cap B)\capdot C=(A\capdot C)\cap (B\capdot C)\,.$$

Multiplicities are taken into account when multisets appear as index
sets (or rather as ``index multisets''). For example, if
$A=\{2,2,3,5,5,5\}$, we have
$$\eqalign{\{\,x-1\mid x\in A\,\}&=\{1,1,2,4,4,4\}\,;\cr
\noalign{\smallskip}
\sum_{x\in A}(x-1)&=\sum\{\,x-1\mid x\in A\}=16\,;\cr
\noalign{\smallskip}
\biguplus_{x\in A}B_x&=B_2\uplus B_2\uplus B_3\uplus B_5\uplus
B_5\uplus B_5\,.\cr}$$
If $P(n)$ is the multiset of prime factors of~$n$, we have 
$\prod\{\,p\mid p\in P(n)\,\}=n$ for all positive integers~$n$. 

If $A$ and $B$ are multisets, we also write
$$\eqalign{A+B&=\{\,a+b\mid a\in A,b\in B\,\}\,,\cr
AB&=\{\,ab\mid a\in A,b\in B\,\}\,;\cr}$$
therefore if $A$ has $m$ elements and $B$ has $n$ elements, both
multisets $A+B$ and $AB$ have $mn$~elements. Notice that
$$\eqalign{[x]\,(A+B)&=\sum_{a\in A}\,[x-a]\,B=\sum_{b\in B}\,[x-b]\,A\cr
\noalign{\smallskip}
&=\sum_{a\in A}\,\sum_{b\in B}\,[x=a+b]\cr}$$
where $[x=a+b]$ is 1 if $x=a+b$ and 0 otherwise. Similar formulas hold
for $[x]\,(AB)$.

It is convenient to let $Ab$ stand for the multiset
$$Ab=\{\,ab\mid a\in A\,\}=A\{b\}\,;$$
similarly, $aB$ stands for $\{a\}B$. This means, for example, that
$2A$ is not the same as $A+A$; a~special notation, perhaps $n\ast A$,
is needed for the multiset
$$\overbrace{A+\cdots +A}^{n\;{\rm times}}=\{\,a_1+\cdots +a_n\mid
a_j\in A\;{\rm for}\;1\leq j\leq n\,\}\,.$$
Similarly we need notations to distinguish the multiset
$$AA=\{\,aa'\mid a,a'\in A\,\}$$
from the quite different multiset
$$\{\,a^2\mid a\in A\,\}=\{\,aa\mid a\in A\,\}\,.$$
The product
$$\overbrace{A\,\ldots\,A}^{n\;{\rm times}}=\{\,a_1\,\ldots\,a_n\mid
a_j\in A\;{\rm for}\;1\leq j\leq n\,\}$$
is traditionally written $A^n$, and I propose writing
$$A\uparrow n=\{\,a^n\mid a\in A\,\}=\{\,a\uparrow n\mid a\in A\,\}$$
on the rarer occasions
when we need to deal with multisets of $n\/$th powers.

\bn{\bf Multilanguages.}\enspace
A {\it multilanguage\/} is like a language, but its elements can
appear more than once. Thus, if we regard a language as a set of
strings, a~multilanguage is a multiset of strings.

An {\it alphabet\/} is a finite set of disinguishable characters. If
$\Sigmait$ is an alphabet, $\Sigmait^{\ast}$~denotes the set
of all strings over~$\Sigmait$. Strings are generally represented by
lowercase Greek letters; the empty string is called~$\epsilon$.
If $A$ is any multilanguage, we write
$$\eqalign{A^0&=\{\epsilon\}\,,\cr
A^{\ast}&=A^0\uplus A^1\uplus A^2\uplus \,\cdots =\biguplus_{n\geq
0}A^n\,;\cr}$$ 
this will be a language (i.e., a~set) if and only if the string
equation $\alpha_1\ldots\alpha_m=\alpha'_1\ldots\alpha'_{m'}$ for
$\alpha_1,\ldots,\alpha_m,\alpha'_1,\ldots,\alpha'_{m'}\in A$ implies
that $m=m'$ and that $\alpha_k=\alpha'_k$ for $1\leq k\leq m$. If
$\epsilon\notin A$, every element of~$A^{\ast}$ has finite
multiplicity; otherwise every element of~$A^{\ast}$ has infinite
multiplicity.

A {\it context-free grammar\/} ${\cal G}$ has four component parts
$(T,N,S,{\cal P})$: $T$~is an alphabet of {\it terminals\/}; $N$~is an
alphabet of {\it nonterminals}, disjoint from~$T$; $S$~is a finite
multiset of {\it starting strings\/} over the alphabet $V=T\cup N$;
and ${\cal P}$ is a finite multiset of {\it productions}, where each
production has the form
$$A\ra\theta\,,\quad\hbox{for some $A\in N$ and $\theta\in
V^{\ast}$}.$$ 
We usually use lowercase letters to represent elements of~$T$, upper
case letters to represent elements of~$N$. The starting strings and
the righthand sides of all productions are called the {\it basic
strings\/} of~${\cal G}$. The multiset $\{\,\theta\mid A\ra
\theta\in {\cal P}\,\}$ is denoted by ${\cal P}(A)$; thus we can
regard ${\cal P}$ as a mapping from~$N$ to multisets of strings
over~$V$.

The productions are extended to relations between strings in the usual
way. Namely, if $A\ra\theta$ is in~${\cal P}$, we say that
$\alpha A\omega$ produces $\alpha\theta\omega$ for all
strings~$\alpha$ and~$\omega$ in~$V^{\ast}$; in symbols, $\alpha
A\omega\ra\alpha\theta\omega$. We also write
$\sigma\ra^n\tau$ if $\sigma$ produces~$\tau$ in $n$~steps;
this means that there are strings $\sigma_0,\sigma_1,\ldots,\sigma_n$
in~$V^{\ast}$ such that $\sigma_0=\sigma$, 
$\sigma_{j-1}\ra\sigma_j$ for $1\leq j\leq n$, and $\sigma_n=\tau$.
 Furthermore we
write $\sigma\ra^{\ast}\tau$ if $\sigma\ra^n\tau$ for
some $n\geq 0$, and $\sigma\ra^+\tau$ if
$\sigma\ra^n\tau$ for some $n\geq 1$.

A {\it parse\/} $\Piit$ for ${\cal G}$ is an ordered forest in which
each node is labeled with a symbol of~$V$; each internal (non-leaf)
node is also labeled with a production of~${\cal P}$. An internal node
whose production label is $A\ra v_1\ldots v_l$ must be labeled with
the symbol~$A$, and it must have exactly $l$~children labeled
$v_1,\ldots,v_l$, respectively. If the labels of the root nodes form
the string~$\sigma$ and the labels of the leaf nodes form the
string~$\tau$, and if there are $n$~internal nodes, we say that
$\Piit$ parses $\tau$ as $\sigma$ in $n$~steps. 
There is an $n$-step parse of~$\tau$ as $\sigma$ if and only
if $\sigma\ra^n\tau$.

In many applications, we are interested in the number of parses; 
so we let $L(\sigma)$ be the multiset of all strings $\tau\in
T^{\ast}$ such that $\sigma\ra^{\ast}\tau$, with each $\tau$
occurring exactly as often as there are parses of~$\tau$ as~$\sigma$.
This defines a multilanguage $L(\sigma)$ for each $\sigma\in
V^{\ast}$.

It is not difficult to see that the multilanguages $L(\sigma)$ are
characterized by the following multiset equations:
$$\eqalign{L(\tau)&=\{\tau\}\,,\quad\hbox{for all $\tau\in
T^{\ast}$}\,;\cr
L(A)&=\biguplus\{\,L(\theta)\mid\theta\in{\cal
P}(A)\,\}\,,\quad\hbox{for all $A\in N$}\,;\cr
L(\sigma\sigma')&=L(\sigma)L(\sigma')\,,\quad\hbox{for all
$\sigma,\sigma'\in V^{\ast}$}\,.\cr}$$
According to the conventions outlined above, the stated formula for $L(A)$
takes account of multiplicities, if any productions
$A\ra\theta$ are repeated in ${\cal P}$. Parse trees that use
different copies of the same production are considered different; we
can, for example, assign a unique number to each production, and use
that number as the production label on internal nodes of the parse.

Notice that the multiplicity of $\tau$ in $L(\sigma)$ is the number of
parses of~$\tau$ as~$\sigma$, not the number of derivations
$\sigma=\sigma_0\ra\cdots\ra\sigma_n=\tau$. For
example, if ${\cal P}$ contains just two productions $\{A\ra a,$ 
{}$B\ra b\}$, then $L(AB)=\{ab\}$ corresponds to the unique
parse
$$\vcenter{\halign{\hfil$#$\hfil\qquad&\hfil$#$\hfil\cr
A&B\cr
\vert&\vert\cr
a&b\cr}}$$
although there are two derivation $AB\ra Ab\ra ab$
and $AB\ra aB\ra ab$. 

The multilanguages $L(\sigma)$ depend only on the alphabets $T\cup N$
and the productions~${\cal P}$. The {\it multilanguage defined
by\/}~${\cal G}$, denoted by $L({\cal G})$, is the multiset of strings
parsable from the starting strings~$S$, counting multiplicity:
$$L({\cal G})=\biguplus\{\,L(\sigma)\mid \sigma\in S\,\}\,.$$

\bn{\bf Transformations.}\enspace
Programmers are especially interested in the way
$L({\cal G})$ changes when ${\cal G}$ is modified. For example, we
often want to simplify grammars or put them into standard forms
without changing the strings of $L({\cal G})$ or their multiplicities.

A nonterminal symbol~$A$ is {\it useless\/} if it never occurs in any
parses of strings in $L({\cal G})$. This happens iff either
$L(A)=\emptyset$ or there are no strings $\sigma\in S$, $\alpha\in
V^{\ast}$, and $\omega\in V^{\ast}$ such that
$\sigma\ra^{\ast} \alpha A\omega$. We can remove all
productions of~${\cal P}$ and all strings of~$S$ that contain useless
nonterminals, without changing $L({\cal G})$. A~grammar is said to be
{\it reduced\/} if every element of~$N$ is useful.

Several basic transformations can be applied to any grammar without
affecting the multilanguage $L({\cal G})$. One of these
transformations is called {\it abbreviation\/}: 
Let $X$ be a new symbol $\notin V$ and let $\theta$ be any string
of~$V^{\ast}$. Add $X$ to $N$ and add the production
$X\ra\theta$ to~${\cal P}$. Then we can replace $\theta$
by~$X$ wherever $\theta$ occurs as a substring of a basic string,
except in the production $X\ra\theta$ itself, without changing
$L({\cal G})$; this follows from the fact that $L(X)=L(\theta)$. By
repeated use of abbreviations we can obtain an equivalent grammar
whose basic strings all have length~2 or less. The total length of all
basic strings in the new grammar is less than twice the total length
of all basic strings in the original.

Another simple transformation, sort of an inverse to abbreviation, is
called {\it expansion}. It replaces any basic string of the form
$\alpha X\omega$ by the multiset of all strings $\alpha\theta\omega$
where $X\ra\theta$. If $\alpha X\omega$ is the right-hand side
of some production $A\ra\alpha X\omega$, this means that the
production is replaced in~${\cal P}$ by the multiset of productions
$\{\,A\ra\alpha\theta\omega\mid\theta\in{\cal P}(X)\,\}$; we
are essentially replacing the element $\alpha X\omega$ of ${\cal
P}(A)$ by the multiset $\{\,\alpha\theta\omega\mid\theta\in{\cal
P}(X)\,\}$. 
Again, $L({\cal G})$ is not affected.

Expansion can cause some productions and/or starting strings to be
repeated. If we had defined context-free grammars differently, taking
$S$ and~${\cal P}$ to be sets instead of multisets, we would not be
able to apply the expansion process in general without losing track of
some parses.

The third basic transformation, called {\it elimination}, deletes a
given production $A\ra\theta$ from~${\cal P}$ and replaces
every remaining basic string~$\sigma$ by $D(\sigma)$, where
$D(\sigma)$ is a multiset defined recursively as follows:
$$\eqalign{D(A)&=\{A,\theta\}\,;\cr
D(\sigma)&=\{\sigma\}\,,\hbox{ if $\sigma$ does not include $A$}\,;\cr
D(\sigma\sigma')&=D(\sigma)D(\sigma')\,.\cr}$$
If $\sigma$ has $n$ occurrences of $A$, these equations imply that
$D(\sigma)$ has $2^n$~elements. Elimination preserves $L({\cal G})$
because it simply removes all uses of the production $A\ra\theta$ from
parse trees.

We can use elimination to make the grammar ``$\epsilon$-free,'' i.e., to
remove all productions whose right-hand side is empty. Complications
arise, however, when a grammar is also ``circular''; this means that
it contains a nonterminal~$A$ such that $A\ra^+A$. The grammars of
most practical interest are non-circular, but we need to deal with
circularity if we want to have a complete theory. It is easy to see
that strings of infinite multiplicity occur in the multilanguage
$L({\cal G})$ of a reduced grammar~${\cal G}$ if and only if ${\cal
G}$ is circular.

One way to deal with the problem of circularity is to modify the
grammar so that all the circularity is localized. Let $N=N_i\cup N_n$,
where the nonterminals of~$N_c$ are circular and those of~$N_n$ are
not. We will construct a new grammar ${\cal G}'=(T,N',S'\cup S'',{\cal
P}')$ with $L({\cal G}')=L({\cal G})$, for which all strings of the
multilanguage $L(S')=\biguplus\{\,L(\sigma)\mid\sigma\in S'\,\}$ have
infinite multiplicity and all strings of
$L(S'')=\biguplus\{\,L(\sigma)\mid\sigma\in S''\,\}$ have finite
multiplicity. The nonterminals of~${\cal G}'$ are $N'=N_c\cup N_n\cup
N'_n\cup N_n''$, where $N'_n=\{\,A'\mid A\in N_n\,\}$ and
$N_n''=\{\,A''\mid A\in N_n\,\}$ are new nonterminal alphabets in
one-to-one correspondence with~$N_n$. The new grammar will be defined
in such a way that $L(A)=L(A')\uplus L(A'')$, where $L(A')$ contains
only strings of infinite multiplicity and $L(A'')$ contains only
strings of finite multiplicity. For each $\sigma\in S$ we include the
members of~$\sigma'$ in~$S'$ and $\sigma''$ in~$S''$, where $\sigma'$
and $\sigma''$ are multisets of strings defined as follows: If $\sigma$
includes a nonterminal in~$N_c$, then $\sigma'=\{\sigma\}$ and
$\sigma''=\emptyset$. Otherwise suppose $\sigma=\alpha_0
A_1\alpha_1\ldots A_n\alpha_n$, where each $\alpha_k\in T^{\ast}$ and
each $A_k\in N_n$; then
$$\eqalign{\sigma'&=\{\,\alpha_0 A''_1\alpha_1\ldots
A''_{k-1}\alpha_{k-1}A'_k\alpha_kA_{k+1}\ldots A_n\alpha_n\mid 1\leq
k\leq n\,\}\,,\cr
\noalign{\smallskip}
\sigma''&=\{\alpha_1A''_1\alpha_1\ldots A''_n\alpha_n\}\,.\cr}$$
(Intuitively, the leftmost use of a circular nonterminal in a
derivation from~$\sigma'$ will occur in the descendants of~$A'_k$. No
circular nonterminals will appear in derivations from~$\sigma''$.) The
productions~${\cal P}'$ are obtained from~${\cal P}$ by letting
$$\eqalign{{\cal P}'(A')&=\biguplus\{\,\sigma'\mid\sigma\in{\cal
P}(A)\,\}\,,\cr
\noalign{\smallskip}
{\cal P}'(A'')&=\biguplus\{\,\sigma''\mid\sigma\in{\cal
P}(A)\,\}\,.\cr}$$
This completes the construction of ${\cal G}'$. 

We can also add a new nonterminal symbol~$Z$, and two new productions
$$\eqalign{Z&\ra Z\,,\cr
Z&\ra\epsilon\,.\cr}$$
The resulting grammar ${\cal G}''$ with starting strings $ZS'\uplus
S''$ again has $L({\cal G}'')=L({\cal G})$, but now all strings with
infinite multiplicity are derived from~$ZS'$. This implies that we can
remove circularity from all nonterminals except~$Z$, without changing
any multiplicities; then $Z$ will be the only source of infinite
multiplicity.

The details are slightly tricky but not really complicated. Let us
remove accumulated primes from our notation, and work with a
grammar~${\cal G}=(T,N,S,{\cal P})$ having the properties just assumed
for~${\cal G}''$. We want ${\cal G}$ to have only~$Z$ as a circular
nonterminal.
The first step is to remove instances of co-circularity: If ${\cal G}$
contains two nonterminals $A$ and~$B$ such that $A\ra^+ B$ and $B\ra^+
A$, we can replace all occurrences of~$B$ by~$A$ and delete $B$
from~$N$. This leaves $L({\cal G})$ unaffected, because every string
of $L({\cal G})$ that has at least one parse involving~$B$ has
infinitely many parses both before and after the change is made.
Therefore we can assume that ${\cal G}$ is a grammar in which
the relations $A\ra^+B$ and $B\ra^+A$
imply $A=B$. 

Now we can topologically sort the nonterminals into order
$A_0,A_1,\ldots,A_m$ so that $A_i\ra^+A_j$ only if $i\leq j$; let
$A_0=Z$ be the special, circular nonterminal introduced above. The
grammar will be in {\it Chomsky normal form\/} if all productions
except those for~$Z$ have one of the two forms
$$A\ra BC\quad{\rm or}\quad A\ra a\,,$$
where $A,B,C\in N$ and $a\in T$. Assume that this condition
holds for all productions whose left-hand side is $A_l$ for some $l$
strictly greater than a given
 index $k>0$; we will show how to make it hold also
for $l=k$, without changing $L({\cal G})$.

Abbreviations will reduce any productions on the right-hand side to
length~2 or less. Moreover, if $A_k\ra v_1v_2$ for $v_1\in T$, we can
introduce a new abbreviation $A_k\ra Xv_2$,
$X\ra v_1$; a~similar abbreviation applies if $v_2\in T$. Therefore
systematic use of abbreviation will put all productions with~$A_k$ on
the left into Chomsky normal form, except those of the forms $A_k\ra
A_l$ or $A_k\ra\epsilon$. By assumption, we can have $A_k\ra A_l$ only
if $l\geq k$. If $l>k$, the production $A_k\ra A_l$ can be eliminated
by expansion; it is replaced by $A_k\ra\theta$ for all $\theta\in{\cal
P}(A_l)$, and these productions all have the required form. If $l=k$,
the production $A_k\ra A_k$ is redundant and can be dropped; this does
not affect $L({\cal G})$, since every string whose derivation uses
$A_k$ has infinite multiplicity because it is derived from $ZS'$.
Finally, a~production of the form $A_k\ra\epsilon$ can be removed by
 elimination as explained above. This does not lengthen the
right-hand side of any production. But it might add new productions of
the form $A_k\ra A_l$ (which are handled as before) or of the form
$A_j\ra\epsilon$. The latter can occur only if there was a production
$A_j\ra A^n_k$ for some $n\geq 1$; hence 
$A_j\ra^+A_k$ and we must have $j\leq k$. If $j=k$, the new production
$A_k\ra\epsilon$ can simply be dropped, because its presence merely
gives additional parses to strings whose multiplicity is already
infinite.

This construction puts ${\cal G}$ into Chomsky normal form, except
for the special 
productions $Z\ra Z$ and $Z\ra\epsilon$, without changing the
multilanguage $L({\cal G})$. If we want to proceed further, we could
delete the production $Z\ra Z$; this gives a grammar~${\cal G}'$ with
$L({\cal G}')\asymp L({\cal G})$ and no circularity. And we can then
eliminate $Z\ra\epsilon$, obtaining a grammar~${\cal G}''$ in Chomsky
normal form with $L({\cal G}'')=L({\cal G}')$. If ${\cal G}$ itself was
originally noncircular, the special nonterminal $Z$ was always useless
so it need not have been introduced; 
our construction produces Chomsky normal form directly in such cases. 

The construction in the preceding paragraphs can be illustrated by the
following example grammar with terminal alphabet $\{a\}$ nonterminal
alphabet $\{A,B,C\}$, starting set $\{A\}$, and productions
$$A\ra AAa\,,\;A\ra B\,,\;A\ra\epsilon\,,\;B\ra CC\,,\;C\ra
BB\,,\;C\ra\epsilon\,.$$
The nonterminals are $N_n=\{A\}$ and $N_c=\{B,C\}$; so we add
nonterminals $N'_n=\{A'\}$ and $N''_n=\{A''\}$, change the starting
strings to
$$S'=\{A'\}\,,\qquad S''=\{A''\}\,,$$
and add the productions
$$\eqalign{&A'\ra A'\!Aa\,,\;A'\ra A''\!A'a\,,\;A'\ra B\,;\cr
&A''\ra A''\!A''a\,,\;A''\ra\epsilon\,.\cr}$$
Now we introduce $Z$, replace $C$ by $B$, and make the abbreviations
$X\ra AY$, $X'\ra A'y$, $X''\ra A''y$, $y\ra a$. 
The current grammar has terminal alphabet $\{a\}$, nonterminal
alphabet $\{Z,A,A',A'',B,\allowbreak
X,X',X'',Y\}$ in topological order, starting strings
$\{ZA',A''\}$, and productions
$$\eqalign{Z&\ra\{Z,\epsilon\}\,,\cr
A&\ra\{AX,B,\epsilon\}\,,\cr
A'&\ra\{A'X,A''X',B\}\,,\cr
A''&\ra\{A''X'',\epsilon\}\,,\cr
B&\ra\{BB,BB,\epsilon\}\,,\cr}$$
plus those for $X$, $X'$, $X''$, $Y$ already stated. Eliminating the
production $B\ra \epsilon$ yields 
new productions $A\ra\epsilon$, $A'\ra\epsilon$; eliminating
$A''\ra\epsilon$ yields a new starting string~$\epsilon$ and
new productions $A'\ra X'$, $A''\ra X''$, $X''\ra a$. We eventually
reach a near-Chomsky-normal grammar with starting strings
$\{Z,ZA',ZA'',A'',\epsilon\}$ and productions
$$\eqalign{%
Z&\ra \{Z,\epsilon\}\,,\cr
A&\ra\{AX,AY,AY,BB,BB,a,a,a,a\}\,,\cr
A'&\ra\{AY,A'X,A'Y,A''X',BB,BB,a,a,a\}\,,\cr
A''&\ra\{A''X'',A''Y,a\}\,,\cr
B&\ra\{BB,BB\}\,,\cr
X&\ra\{AY,a,a\}\,,\cr
X'&\ra\{A'Y,a\}\,,\cr
X''&\ra\{A''Y,a\}\,,\cr
Y&\ra\{a\}\,.\cr}$$

Once a grammar is in Chomsky normal form, we can go further and
eliminate left-recursion. A~nonterminal symbol~$X$ is called {\it
left-recursive\/} if $X\ra^+X\omega$ for some $\omega\in V^{\ast}$. The
following transformation makes $X$ non-left-recursive without
introducing any additional left-recursive nonterminals: Introduce new
nonterminals $N'=\{\,A'\mid A\in N\,\}$, and new productions
$$\displaylines{\{\,B'\ra CA'\mid A\ra BC\in{\cal P}\,\}\,,\cr
\{\,X\ra aA'\mid A\ra a\in{\cal P}\,\}\,,\cr
X'\ra \epsilon\,,\cr}$$
and delete all the original productions of ${\cal P}(X)$. It is not
difficult to prove that $L({\cal G}')=L({\cal G})$ for the new
grammar~${\cal G}'$, because there is a one-to-one correspondence
between parse trees for the two grammars. The basic idea is to
consider all ``maximal left paths'' of nodes labelled
$A_1,\ldots,A_r$, corresponding to the productions
$$A_1\ra A_2B_1\ra A_3B_2B_1\ra\cdots\ra A_rB_{r-1}B_{r-2}\ldots
B_1\ra aB_{r-1}B_{r-2}\ldots B_1$$
in ${\cal G}$, where $A_1$ labels either the root or the right subtree
of~$A_1$'s parent in a parse for~${\cal G}$. If $X$ occurs as at least
one of the nonterminals $\{A_1,\ldots,A_r\}$, say $A_j=X$ but $A_i\neq
X$ for $i<j$, the corresponding productions of~${\cal G}'$ change the
left path into a right path after branch~$j$:
$$\vcenter{\halign{\hfil$#\;$&$#\;$\hfil&$#$\hfil\cr
A_1\ra\cdots\ra A_jB_{j-1}\ldots B_1&\multispan2{$\ra aA'_rB_{j-1}\ldots
B_1\ra aB_{r-1}A'_{r-1}B_{j-1}\ldots B_1$\hfil}\cr
&\ra\cdots&\ra aB_{r-1}\ldots B_jA'_jB_{j-1}\ldots B_1\cr
&&\ra aB_{r-1}\ldots B_jB_{j-1}\ldots B_1\,.\cr}}$$
The subtrees for $B_1,\ldots,B_{r-1}$ undergo the same reversible
transformation.

Once left recursion is removed, it is a simple matter to put the
grammar into {\it Greibach normal form\/}~[3],
in which all productions can be written
$$A\ra aA_1\ldots A_k\,,\qquad k\geq 0\,,$$
for $a\in T$ and $A,A_1,\ldots, A_k\in N$. First we order the
nonterminals $X_1,\ldots,X_n$ so that $X_i\ra X_jX_k$ only when $i<j$;
then we expand all such productions, for decreasing values of~$i$.

\bn{\bf Transduction.}\enspace
A general class of transformations that change one context-free
language into another was discovered by Ginsburg and Rose 
[2],
and the same ideas carry over to multilanguages. My notes from 1964 use
the word ``juxtamorphism'' for a slightly more general class of
mappings; I~don't remember whether I~coined that term at the time or
found it in the literature. At any rate, I'll try it here again and
see if it proves to be acceptable.

If $F$ is a mapping from strings over~$T$ to multilanguages
over~$T'$, it is often convenient to write $\alpha^F$ instead of
$F(\alpha)$ for the image of~$\alpha$ under~$F$. A~family of such
mappings $F_1,\ldots,F_r$ is said to define a {\it juxtamorphism\/}
if, for all~$j$ and for all nonempty strings $\alpha$ and~$\beta$, the
multilanguage $(\alpha\beta)^{F_j}$ can be expressed as a finite
multiset union of multilanguages having ``bilinear form''
$$\alpha^{F_k}\beta^{F_l}\quad{\rm or}\quad
\beta^{F_k}\alpha^{F_l}\,.$$
The juxtamorphism family is called 
context-free if $a^{F_j}$ and $\epsilon^{F_j}$
are context-free multilanguages for all $a\in T$ and all~$j$.

For example, many mappings satisfy this condition with $r=1$. The
reflection mapping, which takes every string $\alpha=a_1\ldots a_m$
into $\alpha^R=a_m\ldots a_1$, obviously satisfies
$(\alpha\beta)^R=\beta^R\alpha^R$. The composition mapping, which
takes $\alpha=a_1\ldots a_m$ into $\alpha^L=L(a_1)\ldots L(a_m)$ for
any given multilanguages $L(a)$ defined for each $a\in T$, satisfies
$(\alpha\beta)^L=\alpha^L\beta^L$. 

The prefix mapping, which takes $\alpha=a_1\ldots a_m$ into
$\alpha^P=\{\epsilon,a_1,a_1a_2,\ldots,a_1\ldots a_m\}$, is a member
of a juxtamorphism family with $r=3$: It satisfies
$$\eqalign{(\alpha\beta)^P&=\alpha^P\beta^E\uplus
\alpha^I\beta^P\,,\cr
(\alpha\beta)^I&=\alpha^I\beta^I\,,\cr
(\alpha\beta)^E&=\alpha^E\beta^E\,,\cr}$$
where $I$ is the identity and $\alpha^E=\epsilon$ for all~$\alpha$.

Any finite-state transduction, which maps $\alpha=a_1\ldots a_m$ into
$$\alpha^T=\{\,f(q_0,a_1)f(q_1,a_2)\ldots
f(q_{m-1},a_m)f(q_m,\epsilon)\,\mid \,q_j\in g(q_{j-1},a_j)\,\}$$
is a special case of a juxtamorphism. Here $q_0,\ldots,q_m$ are
members of a finite set of states~$Q$, and $g$ is a next-state function
from $Q\times T$ into subsets of~$Q$; the mapping~$f$ takes each
member of $Q\times (T\cup\{\epsilon\})$ into a context-free
multilanguage. The juxtamorphism can be defined as follows: Given
$q,q'\in Q$, let $\alpha^{qq'}$ be $\{\,f(q_0,a_1)\ldots
f(q_{m-1},a_m)\mid q_0=q\;{\rm and}\; q_j\in g(q_{j-1},q_j)\;{\rm
and}\; q_m=q'\,\}$. Also let $\alpha^q$ be $\alpha^T$ as described
above, when $q_0=q$. Then
$$\eqalign{(\alpha\beta)^{qq'}&=\biguplus_{q''\in
Q}\alpha^{qq''}\beta^{q''q'}\,;\cr
\noalign{\smallskip}
(\alpha\beta)^q&=\biguplus_{q'\in Q}\alpha^{qq'}\beta^{q'}\,.\cr}$$

The following extension of the construction by Ginsburg and Rose
yields a context-free grammar~${\cal G}_j$ for $L({\cal G})^{F_j}$,
given any juxtamorphism family $F_1,\ldots,F_r$. The grammar~${\cal
G}$ can be assumed in Chomsky normal form, except for a special
nonterminal~$Z$ as mentioned above. The given context-free
multilanguages~$a^{F_j}$ and~$\epsilon^{F_j}$ have terminal
alphabet~$T'$, disjoint nonterminal alphabets $N^{(a,F_j)}$ and
$N^{(\epsilon,F_j)}$, starting strings $S^{(c,F_j)}$ and
$S^{(\epsilon,F_j)}$, productions ${\cal P}^{(a,F_j)}$ and ${\cal
P}^{(\epsilon,F_j)}$. Each grammar~${\cal G}_j$ has all these plus
nonterminal symbols~$A^{F_j}$ for all~$j$ and for all nonterminal~$A$
in~${\cal G}$. Each production $A\ra a$ in~${\cal G}$ leads to
productions $A^{F_j}\ra\{\,\sigma\mid\sigma\in S^{(a,F_j)}\,\}$ for
all~$j$. Each production $A\ra BC$ in~${\cal G}$ leads to the
productions for each~$A^{F_j}$ based on its juxtamorphism
representation. For example, in the case of prefix mapping above we
would have the productions
$$A^P\ra B^PC^E\,,\quad A^P\ra B^IC^P\,,\quad A^I\ra B^IC^I\,,
\quad A^E\ra B^EC^E\,.$$
The starting strings for~${\cal G}_j$ are obtained from those
of~${\cal G}$ in a similar way. 
Further details are left to the reader.

In particular, one special case of finite-state transduction maps
$\alpha$ into $\{k\cdot\alpha\}$ if $\alpha$ is accepted in exactly
$k$~ways by a finite-state automaton. (Let $f(q,a)=a$, and let
$f(q,\epsilon)=\{\epsilon\}$ or~$\emptyset$ according as $q$ is an
accepting state or not.) 
The construction above shows that if $L_1$ is a context-free
multilanguage and $L_2$ is a regular multilanguage, the multilanguage
$L_1\capdot L_2$ is context-free.

\bn{\bf Quantitative considerations.}\enspace
Since multisets carry more information than the underlying sets, we
can expect that more computation will be needed in order to keep track
of everything. From a worst-case standpoint, this is bad news. For
example, consider the comparatively innocuous productions
$$\eqalign{&A_0\ra\epsilon\,,\quad A_0\ra\epsilon\,,\cr
&A_1\ra A_0A_0\,,\quad A_2\ra A_1A_1\,,\quad \ldots\,,\quad 
A_n\ra A_{n-1}A_{n-1}\,,\cr}$$
with starting string $\{A_n\}$. This grammar is almost in Chomsky
normal form, except for the elimination of~$\epsilon$. But
$\epsilon$-removal is rather horrible: There are $2^{2^k}$~ways to
derive~$\epsilon$ from~$A_k$. Hence we will have to replace the
multiset of starting strings by $\{2^{2^n}\cdot\epsilon\}$.

Let us add further productions $A_k\ra a_k$ to the grammar above, for
$0\leq k\leq n$, and then reduce to Chomsky normal form by ``simply''
removing the two productions $A_0\ra\epsilon$. The normal-form
productions will be
$$A_k\ra\left\{\,2^{2^k-2^j+k-j}\cdot A_{j-1}A_{j-1}\mid 1\leq j\leq
k\,\right\}
\biguplus\left\{\,2^{2^k-2^j+k-j}\cdot a_j\mid 0\leq j\leq k\,\right\}\,.$$
Evidently if we wish to implement the algorithms for normal forms, we
should represent multisets of strings by counting multiplicities in
binary rather than unary; even so, the results might blow up
exponentially. 

Fortunately this is not a serious problem in practice, since most
artificial languages have unambiguous or nearly unambiguous grammars;
multiplicities of reasonable grammars tend to be low. And we can at
least prove that the general situation cannot get much worse than the
behavior of the example above: Consider a noncircular grammar with
$n$~nonterminals and with $m$~productions having one of the four forms
$A\ra BC$, $A\ra B$, $A\ra a$, $A\ra \epsilon$. Then the process of
conversion to Chomsky normal form does not increase the set of
distinct right-hand sides~$\{BC\}$ or~$\{a\}$; hence the total number
of distinct productions will be at most $O(mn)$. The multiplicities of
productions will be bounded by the number of ways to attach labels
$\{1,\ldots,m\}$ to the nodes of the complete binary tree with
$2^{n-1}$~leaves, namely~$m^{2^n-1}$.

\bn{\bf Conclusions.}\enspace
String coefficients that correspond to the exact number of parses are
important in applications of context-free grammars, so it is desirable
to keep track of such multiplicities as the theory is developed. This
is nothing new when context-free multilanguages are considered as
algebraic power series in noncommuting variables, except in cases
where the coefficients are infinite. But the intuition that comes from
manipulations on trees, grammars, and automata nicely complements the
purely algebraic approaches to this theory. 
It's a beautiful theory that deserves to be remembered by computer
scientists of the future, even though it is no longer a principal
focus of contemporary research.

Let me close by stating a small puzzle. Context-free multilanguages
are obviously closed under~$\uplus$. But they are not closed
under~$\cup$, because for example the language 
$$\{\,a^ib^jc^id^k\mid
i,j,k\geq 1\,\}\cup\{\,a^ib^jc^kd^j\mid i,j,k\geq 1\,\}$$ 
is inherently ambiguous~[9].
Is it true that $L_1\cup L_2$ is a context-free multilanguage whenever
$L_1$ is context-free and $L_2$ is regular?

\bigskip
\centerline{\bf References}
\bigskip
\disleft20pt:
[1]:
Seymour Ginsburg, {\sl The Mathematical Theory of Context-Free
Languages\/} (New York: Mc\-Graw-Hill, 1966).

\medskip
\disleft20pt:
[2]:
Seymour Ginsburg and G. F. Rose,
``Operations which preserve definability in languages,''
{\sl Journal of the ACM\/ \bf 10} (1963), 175--195.

\medskip\disleft20pt:
[3]:
Sheila A. Greibach, 
``A~new normal-form theorem for context-free pharase structure
grammars,''
{\sl Journal of the ACM\/ \bf 12} (1965), 42--52.

\medskip\disleft20pt:
[4]:
Donald E. Knuth,
``On the translation of languages from left to right,''
{\sl Information and Control\/ \bf 8} (1965), 607--639.

\medskip\disleft20pt:
[5]:
Donald E. Knuth,
``A characterization of parenthesis languages,''
  {\sl Information and Control\/ \bf 11} (1967), 269--289.  

\medskip\disleft20pt:
[6]:
Donald E. Knuth,
``Semantics of context-free languages,''  {\sl Mathematical
 Systems Theory\/ \bf 2} (1968), 127--145.  Errata, 
{\sl Mathematical Systems Theory\/ \bf 5} (1971), 95--96.  

\medskip\disleft20pt:
[7]:
Donald E. Knuth,
{\sl The Art of Computer Programming}, Vol.~2: {\sl Seminumerical Algorithms}
(Reading, Mass.: Addison-Wesley, 1969).

\medskip\disleft20pt:
[8]:
Donald E. Knuth,
``Top-down syntax analysis,''  {\sl Acta Informatica\/ \bf 1}
 (1971), 79--110. 

\medskip\disleft20pt:
[9]:
Rohit J. Parikh,
``On context-free languages,''
{\sl Journal of the ACM\/ \bf 13} (1966), 570--581.

\bye